\documentclass[a4paper, superscriptaddress, amsfonts, amssymb, amsmath, reprint, showkeys, nofootinbib, twoside,aps,prl]{revtex4-1}
\usepackage[english]{babel}
\usepackage[utf8]{inputenc}
\usepackage{xcolor}
\usepackage{bm}
\usepackage{graphicx}
\usepackage{float}
\usepackage{amsmath}
\usepackage{amsfonts,amssymb,amsthm}
\usepackage[colorinlistoftodos, color=green!40, prependcaption]{todonotes}
\usepackage{amsthm}
\usepackage{mathtools}
\usepackage{physics}
\usepackage{xcolor}
\usepackage{graphicx}
\usepackage[left=23mm,right=13mm,top=35mm,columnsep=15pt]{geometry} 
\usepackage{adjustbox}
\usepackage{placeins}
\usepackage[T1]{fontenc}
\usepackage{lipsum}
\usepackage{csquotes}
\usepackage[pdftex, pdftitle={Article}, pdfauthor={Author}]{hyperref} 

\usepackage{geometry}
\begin{document}
\title{A Method for the Time-Frequency Analysis of High-Order Interactions in Non-Stationary Physiological Networks}

\author{Yuri Antonacci}
    \email[Correspondence email address: ]{yuri.antonacci@unipa.it}
    \affiliation{Department of Engineering, University of Palermo, Palermo, Italy}
\author{Chiara Barà}
    \affiliation{Department of Engineering, University of Palermo, Palermo, Italy}
\author{Laura Sparacino}
    \affiliation{Department of Engineering, University of Palermo, Palermo, Italy}
\author{Gorana Mijatovic}
    \affiliation{Faculty of Technical Sciences, University of Novi Sad, Serbia}
\author{Ludovico Minati}
    \affiliation{School of Life Science and Technology, University of Electronic Science and Technology of China, Chengdu, China}
    \affiliation{Center for Mind/Brain Sciences (CIMeC), University of Trento, Trento,}
\author{Luca Faes}
    \affiliation{Department of Engineering, University of Palermo, Palermo, Italy}
    \affiliation{Faculty of Technical Sciences, University of Novi Sad, Serbia}

\begin{abstract}
\textit{Objective}: Several data-driven approaches based on information theory have been proposed for analyzing high-order interactions involving three or more components of a network system. Most of these methods are defined only in the time domain and rely on the assumption of stationarity in the underlying dynamics, making them inherently unable to detect frequency-specific behaviors and track transient functional links in physiological networks. \textit{Methods}: This study introduces a new framework which enables the time-varying and time-frequency analysis of high-order interactions in network of random processes through the spectral representation of vector autoregressive models. The time- and frequency-resolved analysis of synergistic and redundant interactions among groups of processes is ensured by a robust identification procedure based on a recursive least squares estimator with a forgetting factor. \textit{Results}: Validation on simulated networks illustrates how the time-frequency analysis is able to highlight transient synergistic behaviors emerging in specific frequency bands which cannot be detected by time-domain stationary analyses. The application on brain evoked potentials in rats elicits the presence of redundant information timed with whisker stimulation and mostly occurring in the contralateral hemisphere. 
\textit{Conclusions and Significance}: The proposed framework enables a comprehensive time-varying and time-frequency analysis of the hierarchical organization of dynamic networks. As our approach goes beyond pairwise interactions, it is well suited for the study of transient high-order behaviors arising during state transitions in many network systems commonly studied in physiology, neuroscience and other fields.
\end{abstract}


\maketitle

\section{Introduction}
Network models represented as graphs are essential tools for exploring the structure and dynamics of various physiological systems and their interactions \cite{santoro2024higher}. In these models, each entity, such as a brain unit or an organ system, is represented as a node, while the edges map functional dependencies like brain connectivity \cite{rubinov2010complex} or cardiovascular interactions \cite{lehnertz2020human}. Although widely recognized and applied as a standard tool for many complex physiological systems, graph-based networks are limited by the underlying assumption that interactions between nodes are strictly pairwise. However, increasing evidence in neuroscience and physiology underscores the importance of group interactions, which go beyond pairwise connections to involve the collective dynamics of groups of nodes \cite{battiston2020networks}. These interactions are characterized by non-factorizability, meaning that the joint behavior of multiple components cannot be expressed as the sum of individual or pairwise contributions \cite{antonacci2021measuring,battiston2020networks}. Several studies have demonstrated the presence of such non-factorizable interactions, denoted as high-order interactions (HOIs), in physiological systems. For example, HOIs have been identified in cardiovascular and respiratory networks, whose joint dynamics are modulated collectively by multiple coexisting physiological mechanisms \cite{faes2022new,sparacino2024measuring,mijatovic2024assessing,mijatovic2024network}, or in brain networks, where the complex interplay among three or more brain regions has been shown to be crucial for neural processing and behavior \cite{luppi2024unravelling,AntonacciSpectral24,antonacci2021measuring,pirovano2023rehabilitation,sparacino2023statistical}.  

Despite their importance, identifying HOIs in physiological systems remains a significant challenge, because in these systems interactions are not inherently defined but need to be inferred from data \cite{battiston2020networks}. To tackle this issue, various information-theoretic metrics have been proposed, which assess the statistical concepts of synergy and redundancy among groups of random variables or processes \cite{williams2010nonnegative,antonacci2021measuring,faes2022new,mijatovic2024assessing,stramaglia2021quantifying,stramaglia2024disentangling,rosas2019quantifying}. Synergy reflects statistical interactions that uniquely arise from the collective behavior of a network, meaning they cannot be deduced by examining individual components or smaller subsets in isolation; in contrast, redundancy refers to overlapping contributions, where shared information among subgroups of variables sufficiently explains the observed interactions. Interaction Information \cite{mcgill1954multivariate} (II) was one of the first measures proposed to detect the balance between synergy and redundancy. More recently, this concept has been extended through the O-Information \cite{rosas2019quantifying} (OI), which provides a framework to distinguish synergy-dominated from redundancy-dominated interactions in networks of multiple interacting variables. Furthermore, given that physiological networks display temporally correlated activities, dynamic approaches measuring information rates—such as the II rate \cite{faes2022new,antonacci2021measuring} and the OI rate \cite{AntonacciSpectral24,sparacino2024measuring}—have been introduced to replace traditional static quantities. These important developments opened the way to the analysis of HOIs in physiological systems represented as networks of random processes.

One of the major challenges in studying dynamic network systems is the prevalence of non-stationary behaviors \cite{hesse2003use,ivanov2021new,antonacci2023time,antonacci2024exploring}. In particular, the output dynamics of physiological systems transiently change over time with different physiological states, pathological conditions, or external stimuli \cite{ivanov1998stochastic}, directly affecting pairwise and higher-order behaviors. Therefore, there is a critical need for methods capable of tracking HOIs in a time-resolved fashion. Moreover, since physiological systems typically exhibit oscillatory behaviors deployed across distinct frequency bands \cite{sparacino2024measuring}, the development of spectral approaches is also envisaged. To address these needs, the present study introduces a framework for analyzing redundant and synergistic HOIs in multivariate stochastic processes, capturing their time-varying and time-frequency dynamics. The framework leverages the time-resolved entropy rate \cite{barbieri2004dynamic,antonacci2023time} and its frequency-domain expansion \cite{sparacino2024method} as main tools to characterize HOIs in nonstationary multivariate processes with coupled oscillatory components. This is achieved using linear time-varying vector autoregressive (VAR) models \cite{haykin2002adaptive}, whose frequency domain representation is exploited to assess HOIs at each time step. Model identification is performed through the recursive least squares (RLS) algorithm, which has been successfully applied to map the information content of brain signals \cite{antonacci2023time,antonacci2024exploring} and their interactions, even when only a single realization of the process is available \cite{moller2001instantaneous,moller2003fitting,milde2010new,astolfi2008tracking,cerutti2002time,plomp2014physiological,pagnotta2018time}.

The proposed framework is first illustrated through theoretical examples of simulated VAR processes featuring time-variant HOIs of different types and orders across various frequency bands. It is then applied to study the processing of evoked potentials in electroencephalographic (EEG) brain signals acquired from rats during whisker stimulation \cite{plomp2014physiological}, where HOIs are expected to play a crucial role in governing collective dynamics.

\section{Methods}

\subsection{Time-resolved O-information rate}

Let us consider a set of $M$ signals, e.g. measuring the activity of different brain regions, as a realization of the multivariate (vector) stochastic process $Y=\{Y_1,\ldots,Y_M\}$. Let us also denote as $Y(t_n) = [Y_{1}(t_n),\ldots,Y_{M}(t_n)]^\top \in \mathbb{R}^{M\times 1}$ the vector random variable that samples the process at the temporal index $n\in \mathbb{N}$ ($t_n=n\Delta t$, $\Delta t=1/f_s$, $f_s$ sampling frequency). 
Then, considering a generic subset $X\subseteq Y$ containing $N\leq M$
processes, $X=\{X_1,\ldots,X_N\}=\{Y_{i_1},\ldots,Y_{i_N}\}$, $i_k \in
\{1,\ldots,M\}$, $k=1,\ldots, N$, the average rate of information produced at the time $t_n$ by $X$ is measured through its \textit{entropy rate} (ER) \cite{barbieri2004dynamic}: 
\begin{equation}
H_{X}(t_n)=H(X(t_n)|X^q(t_n)),  
\label{eq:ER}
\end{equation}
where $X(t_n) \in \mathbb{R}^{N \times 1}$ is the variable sampling $X$ at the present time $t_n$,  $X^q(t_n)=[X(t_{n-1}),\ldots,X(t_{n-q})]^\top \in \mathbb{R}^{q N \times 1}$ collects the variables sampling $X$ over $q$ past lags (in theory, $q \rightarrow \infty$), and $H(\cdot|\cdot)$ denotes conditional entropy \cite{cover1999elements}.
The ER (\ref{eq:ER}) measures the complexity of the process $X$ at a specific time point, capturing the amount of information contained in the present time of $X$ that cannot be explained by its past history. This measures ranges from $H_{X}(t_n)=0$ for a fully self-predictable process to $H_{X}(t_n)=H(X(t_n))$ for a fully unpredictable process \cite{mijatovic2024assessing,AntonacciSpectral24}. 

Starting from the time-specific ER (\ref{eq:ER}), we formulate a time-specific HOI measure which generalizes to non-stationary random processes the so-called O-information rate (OIR) defined in \cite{faes2022new,sparacino2024measuring}. Specifically, the time-resolved OIR of the vector process $X=\{X_1,\ldots,X_N\}$ is defined as:
\begin{equation}
    \Omega_{X}(t_n)=(N-2)H_{X}(t_n)+\sum_{j=1}^N \left[H_{X_j}(t_n)-H_{X^{j}}(t_n)\right] \label{tvOIR},
\end{equation}
where $X^{j}=X\setminus X_j$ is the vector process obtained removing $X_j$ from $X$.
Note that for stationary random processes the time-resolved OIR is the same at each time step $t_n$, $\Omega_{X}(t_n)=\Omega_{X}$, returning the time-invariant OIR measure \cite{faes2022new}. 
Eq. (\ref{tvOIR}) provides a null value when $N=2$ processes are analyzed, while in the case $N=3$ it yields a time-resolved extension of the well-known interaction information rate (IIR) \cite{faes2022new,mcgill1954multivariate,antonacci2021measuring}.
Crucially, the time-specific OIR can be positive or negative, with the sign reflecting the redundant or synergistic nature of the analyzed group of processes. Specifically, $\Omega_X(t_n)$ takes positive values when redundant interactions dominate over synergistic ones and negative values when the opposite occurs.  

\subsection{Computation for linear multivariate processes} \label{TIME-RES-OIR}

Under the assumption that the variables sampling the original vector process at each time $t_n$ have a joint Gaussian distribution, the time-specific OIR (\ref{tvOIR}) can be computed in the framework of VAR modeling \cite{faes2016information}. Specifically, the current state of the overall process $Y$ can be represented by a linear combination of the past states by means of a linear time-varying VAR (TV-VAR) model \cite{haykin2002adaptive}:
\begin{equation}
    Y(t_n)=\sum_{k=1}^p \mathbf{A}_{k}(t_n)Y(t_{n-k}) +U(t_n),
    \label{TV-VAR}
\end{equation}
where $\mathbf{A}_{k}(t_n) \in \mathbb{R}^{M\times M}$ is the matrix of the model coefficients and $U(t_n)=[U_{1}(t_n),\ldots,U_{M}(t_n)]^\top \in \mathbb{R}^{M\times 1}$ is a vector of zero-mean innovation variables with $M\times M$ covariance matrix $\mathbf{\Sigma}_{U}(t_n)=\mathbb{E}[U(t_n) U(t_n)^\top]$. 
Then, to describe the dynamics relevant to a generic $Q$-dimensional sub-process $S \subseteq X$, $S \in \mathbb{R}^{Q\times 1}$, we exploit a restricted TV-VAR model:
\begin{equation}
    S(t_n)=\sum_{k=1}^q \mathbf{B}_k(t_n)S(t_{n-k})+U_{S}(t_n), \label{restricted}
\end{equation}
where $\mathbf{\Sigma}_{U_{S}}(t_n) \in \mathbb{R}^{Q \times Q}$ is the covariance matrix of the innovation process feeding the restricted model and $\mathbf{B}_k(t_n) \in \mathbb{R}^{Q \times Q}$ is the relevant coefficient matrix; note that the order $q$ of the restricted model will generally tend to infinity because a sub-process of a VAR process will have a moving average component \cite{faes2017interpretability}. The time-specific ER of the sub-process $S$ can be then derived from the generalized variance of the residuals in the restricted model (\ref{restricted}), as \cite{sparacino2024measuring}: 
\begin{equation}
    H_{S}(t_n)=\dfrac{1}{2}\log{(2 \pi e)^{Q} |\mathbf{\Sigma}_{U_{S}}(t_n)|}, \label{ERres}
\end{equation}
where $|\cdot|$ stands for matrix determinant. Thus, by applying (\ref{ERres}) $2N+1$ times, each with $S$ taking the role of one of the processes appearing in (\ref{tvOIR}), yields to compute the time-specific OIR for the set of processes collected in $X$. 

The parameters of the full model (\ref{TV-VAR}), i.e., $\mathbf{A}_k(t_n)$ and $\mathbf{\Sigma}_{U}(t_n)$, are estimated with the TV-VAR identification procedure described in Sect. \ref{identification}. Then, the parameters of each restricted model (\ref{restricted}), i.e., $\mathbf{B}_k(t_n)$ and $\mathbf{\Sigma}_{U_{S}}(t_n)$, are derived from those of the full model through a two-step procedure which (i) derives the time-lagged covariance structure of the full process at each time step $Y(t_n)$ by solving the Yule-Walker (YW) equations at the relevant time-step, and (ii) reorganizes such structure to relate it to the covariance of the sub-process $S$; a detailed description of the practical solution of the YW equations can be found in \cite{antonacci2023time,mijatovic2024network}. The order $p$ of the full model is estimated as described in Sect. \ref{identification}, while the order $q$ of the restricted models is set at a high value to capture the decay of the correlation function \cite{barnett2014mvgc}.

\subsection{Time-frequency expansion} \label{time-frequency HOIs}
The time-frequency analysis of HOIs can be implemented exploiting the spectral representation of VAR models \cite{faes2012measuring} and observing that the TV-VAR model (\ref{TV-VAR}) admits a stationary representation at each time step $t_n$ \cite{robert1996continuously}. This representation provides the transfer matrix relating the Fourier transform (FT) of the processes $U$ and $Y$ in (\ref{TV-VAR}), obtained at each time instant $t_n$ as the inverse of the FT of the time-specific model coefficients through the relation \cite{hodgkiss1981adaptive,moller2001instantaneous,faes2012measuring}
\begin{equation}
\mathbf{H}(t_n,\omega)=\left[\mathbf{I}-\sum_{k=1}^p \mathbf{A}_{k}(t_n) e^{-\mathbf{j} \omega k}\right]^{-1},
   \label{TransferMatrix}
\end{equation}
where $\omega \in [-\pi,\pi]$ is the normalized angular frequency ($\omega=2\pi \frac{f}{f_s}$ with $f\in \left[-\frac{f_s}{2},\frac{f_s}{2}\right]$), 
$\mathbf{j}=\sqrt{-1}$, and $\mathbf{I}$ is the $M \times M$ identity matrix.

Starting from (\ref{TransferMatrix}), the power spectral density (PSD) of the overall process relevant to the time-step $t_n$ can be expressed as $\mathbf{P}_{Y}(t_n,\omega)=\mathbf{H}(t_n,\omega)\mathbf{\Sigma}_{U}(t_n)\mathbf{H}^*(t_n,\omega)$ \cite{faes2012measuring}, where $^*$ stands for conjugate transpose. Then, the PSD matrix of $Y$ can be pruned to extract the PSD of any generic sub-process $S$ satisfying (\ref{restricted}), $\mathbf{P}_{S}(t_n,\omega)$. Based on this partition, the spectral ER of the sub-process $S$ is defined as \cite{antonacci2021measuring,sparacino2024measuring} 
\begin{equation}
    h_{S}(t_n,\omega)=\dfrac{1}{2}\log{(2 \pi e)^{Q} |\mathbf{P}_{S}(t_n,\omega)|}, \label{spectralER}
\end{equation}
yielding a time-frequency measure of the complexity of the process $S$. Importantly, the spectral and time domain definitions of ER given in (\ref{spectralER}) and (\ref{ERres}) are closely related to each other by the spectral integration property \cite{chicharro2011spectral}, in a way such that the integration of the spectral ER over the whole frequency axis returns the time-specific ER  (\ref{ERres}) \cite{sparacino2024measuring}:
\begin{equation}
H_{S}(t_n)=\dfrac{1}{2\pi} \int_{-\pi}^{\pi} h_{S}(t_n,\omega) \textrm{d}\omega. \label{SpectIntER} 
\end{equation}
Given (\ref{SpectIntER}) and (\ref{tvOIR}) it is easy to show that a spectral version of the time-specific OIR of the $N$ processes in $X$ is given by:
\begin{equation}
     \nu_X(t_n,\omega)=(N-2)h_X(t_n,\omega)+\sum_{j=1}^N \left[h_{X_j}(t_n,\omega)-h_{X^j}(t_n,\omega)\right], 
     \label{spectralOIR}
\end{equation}
which, in turn, satisfies the spectral integration property, i.e., $\Omega_X(t_n)=\frac{1}{2\pi} \int_{-\pi}^{\pi}\nu_X(t_n,\omega) \textrm{d}\omega$.
Therefore, the spectral version of the time-resolved HOI measure (\ref{tvOIR}) defined in (\ref{spectralOIR}) can be meaningfully interpreted as a time- and frequency-specific index of the synergistic/redundant information shared by multiple stochastic processes.

\subsection{Model identification} \label{identification}
Here we describe the procedure followed for the practical computation of the time-frequency OIR measure, which is based on the RLS estimation of the TV-VAR model (\ref{TV-VAR})  \cite{antonacci2023time,moller2001instantaneous}.
The procedure starts with $R$ realizations of the analyzed vector process $Y$, available in the form of $R$ multivariate time series each composed by $M$ series measured over $T$ samples: $y^{(r)}(t_i) \in \mathbb{R}^{M \times 1}, r=1,\ldots,R, i=1,\ldots,T$. From this dataset, the following observation matrices are formed for each time step $t_n$, $n=p+1,\ldots,T$ : $\mathbf{Y}(t_n)=[y^{(1)}(t_n),\ldots, y^{(R)}(t_n)] \in \mathbb{R}^{M \times R}$, containing the observations of the present state of $Y$; $\mathbf{W}(t_n)=[w^{(1)}(t_n),\ldots,w^{(R)}(t_n)]\in \mathbb{R}^{Mp \times R}$, with $w^{(r)}(t_n)=[y^{(r)}(t_{n-1})^\top,\ldots, y^{(r)}(t_{n-p})^\top]^\top \in \mathbb{R}^{Mp \times 1}$, containing the observations of the $p$ past states of $Y$. With this notation, a compact representation of the time-varying model (\ref{TV-VAR}) can be obtained as: $\mathbf{Y}(t_n)=\mathbf{A}(t_n)\mathbf{W}(t_n)+\mathbf{U}(t_n)$, where $\mathbf{A}(t_n)=[\mathbf{A}_{1}(t_n),\ldots,\mathbf{A}_{p}(t_n)] \in \mathbb{R}^{M \times Mp}$ is the matrix of unknown VAR coefficients at time $t_n$ while $\mathbf{U}(t_n)=[u^{(1)}(t_n),\ldots,u^{(R)}(t_n)]\in \mathbb{R}^{M \times R}$, with $u^{(r)}(t_n) \in \mathbb{R}^{M \times 1}$ contains the observations of the innovation process $U$.

The RLS involves the minimization of a cost function defined as \cite{haykin2002adaptive,antonacci2023time}: $\mathcal{E}(\mathbf{Y},\hat{\mathbf{Y}})=\sum_{n=p+1}^{T} (1-c)^{T-n}||\mathbf{Z}(t_n)||^2$, where the matrix $\mathbf{Z}(t_n)=\mathbf{Y}(t_n)-\hat{\mathbf{A}}(t_{n-1})\mathbf{W}(t_n) \in \mathbb{R}^{M \times R}$ denotes the \textit{a-priori} estimation error computed as difference between the matrix of the real data $\mathbf{Y}(t_n)$ and the estimated data $\hat{\mathbf{Y}}(t_n)$. The term $(1-c)^{T-n}$ is the exponential weighting factor or forgetting factor, with $0 < (1-c) \leq 1$, which can be roughly interpreted as the memory of the algorithm ensuring that the data in the distant past are “forgotten” to follow the variations of the statistical properties of $Y$ in non-stationary conditions; this parameter controls the trade-off between the adaption speed at transition and the variance of the estimate. The RLS algorithm to estimate the matrix of VAR coefficients consists in \cite{antonacci2023time,milde2010new}:  (i) choose a value for the adaptation factor $c$ and an order of the TV-VAR model $p$; (ii) define proper initial conditions for the VAR coefficients at time $t_p$, $\mathbf{A}(t_p)=[\mathbf{A}_{1}(t_p),...,\mathbf{A}_{p}(t_p)] \in \mathbb{R}^{M\times Mp}$, and for the correlation matrix of the past states of $Y$ stored in $\mathbf{W}(t_p)$, $\mathbf{\Phi}^{w}(t_p) \in \mathbb{R}^{Mp\times Mp}$; (iii) for $t_{n}=t_{p+1}$ to $t_T$ repeat the following steps:
\begin{subequations}
    \begin{align}
     \mathbf{\Phi}^{w}(t_n)=(1-c)\mathbf{\Phi}^{w}(t_{n-1})+\mathbf{W}(t_n)\mathbf{W}(t_n)^\top, \label{PHI}\\
     \mathbf{K}(t_n)=(\mathbf{\Phi}^{w}(t_n))^{-1}\mathbf{W}(t_n), \label{gain}\\
     \mathbf{Z}(t_n)=\mathbf{Y}(t_n)-\hat{\mathbf{A}}(t_{n-1})\mathbf{W}(t_n), \label{inst-error}\\ 
     \hat{\mathbf{A}}(t_n)=\hat{\mathbf{A}}(t_{n-1})+\mathbf{Z}(t_n) \mathbf{K}(t_n)^\top, \label{varvaring}
    \end{align}
    \label{FULL_RLSID}
\end{subequations}
\noindent{where $\mathbf{K}(t_n) \in \mathbb{R}^{Mp\times R}$ is the so-called gain matrix. A detailed mathematical derivation of the RLS solution is reported in the supplemental material of \cite{antonacci2023time}.}
When $0<c<1$, a recursive computation of the time-varying covariance matrix of the residuals $\mathbf{\Sigma}_{U}(t_n)$ can be obtained as follows \cite{grieszbach1994dynamic}:
\begin{equation}
    \hat{\mathbf{\Sigma}}_{U}(t_n)= \hat{\mathbf{\Sigma}}_{U}(t_{n-1})+c\left(\frac{\mathbf{Z}(t_n) \mathbf{Z}(t_n)^\top}{R} -\hat{\mathbf{\Sigma}}_{U}(t_{n-1})\right), \label{tvar-varU} 
\end{equation}
while when $c=0$ (i.e., assuming stationarity for $Y$):
\begin{equation}
    \hat{\mathbf{\Sigma}}_{U}(t_n)= \hat{\mathbf{\Sigma}}_{U}(t_{n-1})+\dfrac{1}{n}\left(\frac{\mathbf{Z}(t_n) \mathbf{Z}(t_n)^\top}{R} -\hat{\mathbf{\Sigma}}_{U}(t_{n-1})\right).
\end{equation}

The model order of the TV-VAR model (\ref{TV-VAR}) can be determined through computation of the mean squared prediction error (MSPE), $\dfrac{1}{T-p}\sum_{n=p+1}^{T}||\mathbf{Z}(t_n)||^2$; specifically, the optimal order is selected as the minimum MSPE averaged over the $M$ processes after scanning for values of $p$ between 1 and a maximum order $P$\cite{costa2011adaptive,chicharro2011spectral}.
We note that when $c=0$ the process $Y$ is considered as globally stationary and the RLS algorithm is applied "with infinite memory" \cite{haykin2002adaptive} reducing to the well-known ordinary least squares (OLS) estimator \cite{lutkepohl2005new} applied over the whole 
time series of length $T$. On the other hand, by assuming $c=1$ the process is interpreted as "memory-less" and, if multiple realizations are available ($R\geq M^2p$ to ensure the existence of the solution \cite{antonacci2024measuring}), the use of the OLS estimator allows a time-resolved identification; specifically, by setting $c=1$ in (\ref{FULL_RLSID}) the VAR coefficients are estimated at each time step $t_n\geq t_{p+1}$ through the well-known OLS formula: $\hat{\mathbf{A}}(t_n)=\mathbf{Y}(t_n)\mathbf{W}(t_n)^\top [\mathbf{W}(t_n) \mathbf{W}(t_n)^\top]^{-1}$ and
$\hat{\mathbf{U}}(t_n)=\mathbf{Y}(t_n)-\hat{\mathbf{A}}(t_n)\mathbf{W}(t_n)$, whose covariance matrix $\hat{\mathbf{\Sigma}}_{U}(t_n)$ is an estimate of $\mathbf{\Sigma}_{U}(t_n)$. 

\section{Simulation Study}
This section presents the design and implementation of a benchmark simulation of multiple interacting stochastic processes highlighting the peculiar features of the proposed time-varying and time-frequency analysis of HOIs. Specifically, we show how different patterns of redundancy and synergy can emerge in different time windows and coexist at different frequencies, and would  remain undetected if the standard stationary time-domain analysis was performed.
Additionally, we examine the performance of the RLS algorithm in tracking sudden changes in HOIs, emphasizing the role of the forgetting factor and of the number of realizations available for the estimation. The MATLAB code to reproduce the simulation study, along with a detailed documentation, is available at: \url{https://github.com/YuriAntonacci/tv-OIR-toolbox}.

\begin{figure}
\includegraphics{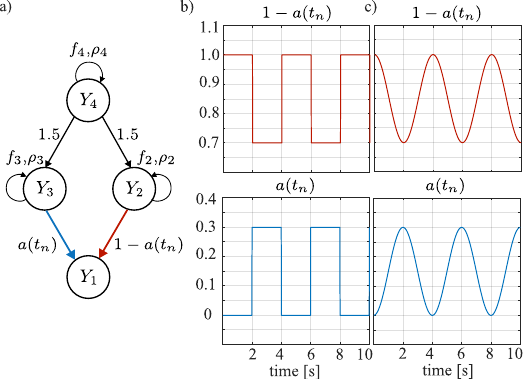} 
    \caption{(a) Graphical representation of the time-varying VAR process generated according to (\ref{SIM}), where each node represents a different scalar process and the arrows represent the imposed causal interactions (self-loops depict the influences from the past to the present sample of a process). The blue and red arrows represent time-varying coupling parameters which are modulated as depicted in (b) and (c).}\label{fig:1}
\end{figure}

\subsection{Theoretical analysis}
\begin{figure*}
\includegraphics{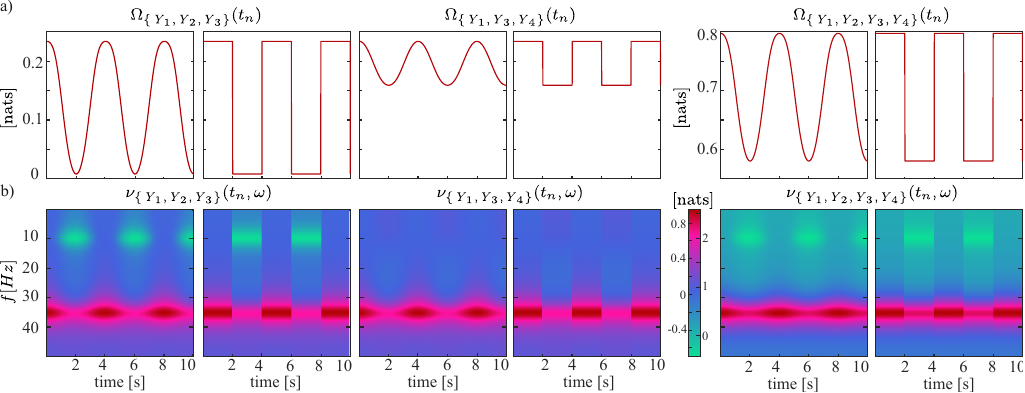}
\caption{Time-resolved OIR and its spectral version computed for the simulated VAR process (\ref{SIM}), driven by square and sinusoidal periodic waveforms oscillating as depicted in Fig. \ref{fig:1}. (a) Time-specific OIR measure of $3^{rd}$- (i.e., $\Omega_{\{Y_1,Y_2,Y_3\}}(t_n)$, $\Omega_{\{Y_1,Y_3,Y_4\}}(t_n)$) and $4^{th}$-order (i.e., $\Omega_{\{Y_1,Y_2,Y_3,Y_4}(t_n)\}$) evidencing the presence of redundancy and multiple state transitions every 2 s. (b) Time-frequency profiles of $3^{rd}$- and $4^{th}$-order highlighting the coexistence of redundant and synergistic effects in different frequency bands.}
\label{fig2}
\end{figure*}
We design a four-variate time-varying VAR process with regression coefficients modulated temporally according to predefined waveforms for inducing multiple transitions within a predefined time window \cite{antonacci2021measuring,antonacci2023time}.
The analyzed TV-VAR process is:
\begin{equation}
\begin{aligned}
Y_{1}(t_n)&=(1-a(t_{n}))Y_{2}(t_{n-1})+a(t_{n})Y_{3}(t_{n-2})+U_{1}(t_n), \\
Y_{2}(t_n)&=\sum_{k=1}^2 a_{2k}Y_{2}(t_{n-k})+1.5Y_{4}(t_{n-1})+U_{2}(t_{n}), \\
Y_{3}(t_n)&=\sum_{k=1}^2 a_{3k}Y_{3}(t_{n-k})+1.5Y_{4}(t_{n-2})+U_{3}(t_{n}), \\
Y_{4}(t_n)&=\sum_{k=1}^2 a_{4k}Y_{4}(t_{n-k})+U_{4}(t_{n}),
\end{aligned}
\label{SIM}
\end{equation}
where $U_{1}(t_{n}),\ldots,U_{4}(t_{n})$ are white and uncorrelated Gaussian processes with zero mean and unit variance.
The process is simulated assuming a sampling frequency $f_s=100$ Hz. The time-varying nature of the process is determined by the coupling parameter $a(t_{n})$ modulating reciprocally the strength of the causal connections $Y_2\rightarrow Y_1$ and $Y_3\rightarrow Y_1$ as reported in Fig. \ref{fig:1};
specifically, $a(t_{n})$ is set to vary over time as a periodic square waveform, or as a periodic sinusoidal waveform, both oscillating in amplitude between 0 and 0.3 with a period of 4 s. The other simulated causal connections are $Y_4 \rightarrow Y_3$ and $Y_4 \rightarrow Y_2$, both with time-independent strength set to 1.5. Moreover,  autonomous oscillations are set for the processes $Y_i$, $i=2,3,4$, by setting time-independent coefficients $a_{i1}=2\rho_i \mathrm{cos}(2\pi f_i/f_s)$ and $a_{i2}=-\rho_i^2$, with $\rho_2=\rho_3=\rho_4=0.85$ and $f_2=f_3=10$ Hz and $f_4=35$ Hz, so as to resemble the $\alpha$ and $\beta$ brain rhythms.

The time-frequency representation of the TV-VAR process can be obtained by deriving its PSD directly from the true TV-VAR coefficients as described in Section \ref{time-frequency HOIs}, from which the exact values of both the spectral and the time-specific versions of the OIR are obtained at any order of interaction. In particular, the theoretical profiles of the time-varying and time-frequency representation of the OIR are shown in Fig. \ref{fig2} for multiplets of order three ($X=\{Y_1,Y_2,Y_3\}$ and $X=\{Y_1,Y_3,Y_4\}$) and for the full process ($X=Y$).
The time-specific analysis (Fig. \ref{fig2}a) reveals the predominance of redundancy in the system, documented by the positive values displayed by the time-domain OIR for all the analyzed multiplets of processes. The amount of net redundancy is modulated over time according to the variable coupling from $Y_3$ and $Y_2$ to $Y_1$, with OIR values changing proportionally to ($1-a(t_n)$). These results document the importance of performing a time-varying analysis in order to track the variations over time of the redundancy/synergy balance, here modulated by the strength of the chain effect $Y_4 \rightarrow Y_2 \rightarrow Y_1$ (producing redundancy) relative to that of the common child effect $Y_3 \rightarrow Y_1 \leftarrow Y_2$ (producing synergy) \cite{antonacci2021measuring,sparacino2024measuring}.
Moreover, the time-frequency analysis of HOIs (Fig. \ref{fig2}b) reveals the coexistence of redundant and synergistic interactions with extent depending on the coupling parameter $a(t_n)$. Specifically, the time-frequency OIR $\nu_{\{Y_1,Y_2,Y_3\}}(t_n,\omega)$, and more smoothly also $\nu_{\{Y_1,Y_2,Y_3,Y_4\}}(t_n, \omega)$, display net redundancy in the band centered around 35 Hz when $a(t_n)\sim0$, which becomes less pronounced and  accompanied by evident net synergy around 10 Hz when $a(t_n)>0$.
These patterns document the importance of analyzing time-varying HOIs in the frequency domain to reveal effects that may remain hidden in the time domain: in our simulation the chain $Y_4 \rightarrow Y_2 \rightarrow Y_1$ and the common child $Y_3 \rightarrow Y_1 \leftarrow Y_2$, which are both active when $a({t_n})$ grows towards 0.3, yield contemporaneously synergy at $\sim 10$ Hz and redundancy at $\sim 35$ Hz, while only the chain effect producing redundancy is present in the time epochs during which $a({t_n})$ takes low values.

\subsection{Practical estimation}

\begin{figure*}
\includegraphics{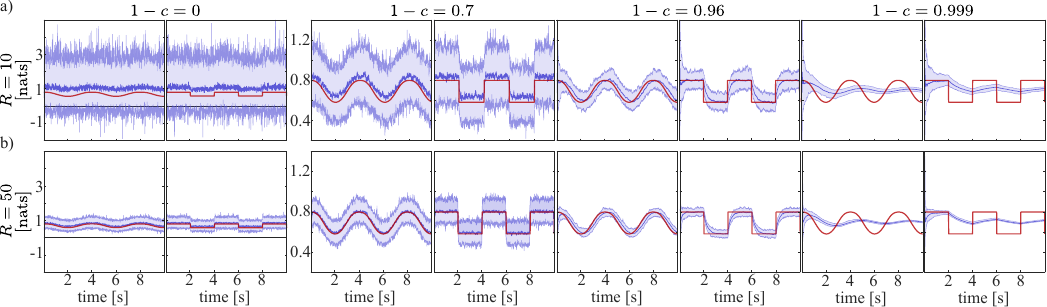}
\caption{Distributions of the OIR $\hat{\Omega}_{\{Y_1,Y_2,Y_3,Y_4\}}(t_n)$ (median and $5^{th}-95^{th}$ percentiles) estimated over 100 iterations of (\ref{SIM}), for different values of $(1-c)\in \{0, 0.7, 0.96, 0.999\}$ and for different waveforms of $a({t_n})$, displayed separately after generating (a) $R=10$ realizations or (b) $R=50$ realizations. In each panel, the profile of the theoretical OIR $\Omega_{\{Y_1,Y_2,Y_3,Y_4\}}(t_n)$ is reported in red.}
\label{fig3}
\end{figure*}

To investigate the performance of the proposed methodology, the time-varying OIR relevant to the whole system (\ref{SIM}), i.e., $\Omega_{X}$ computed with $X=\{Y_1,Y_2,Y_3,Y_4 \}$, was estimated applying the procedure described in Sect. \ref{identification} on realizations of the TV-VAR model of length $T=1000$ (corresponding to a time window of 10 s). Two analyses were performed, both varying the number of realizations generated ($R\in \{10, 20, 50, 100\}$) and the forgetting factor ($(1-c)\in \{0, 0.5, 0.7, 0.9, 0.94, 0.96, 0.97, 0.98, 0.999\}$): the first aimed at studying the capability of the time-varying OIR estimator to follow fast and slow variations of the coupling parameter (square and sinusoidal waveforms), and the second aimed at quantifying bias and variance of the OIR estimated during the stationary phases of the periodic square variations of the coupling parameter.
In both cases, the model order was estimated as described in Sect. \ref{identification}, $q$ was set to 30 and the matrix $\mathbf{A}(t_p)$ was  assigned drawing random samples from the uniform distribution between -1 and 1 \cite{antonacci2024exploring}. 
In the second analysis, for  each combination of the parameters $((1-c),R)$, the normalized bias was computed as $\mathrm{BIAS_N}=\frac{1}{T}\sum_{i=1}^T\frac{|\Omega_{X}(t_i)-\hat{\Omega}_{X}(t_i)|}{\Omega_{X}(t_i)}$, and the variance was computed as 
$\mathrm{VAR}=\frac{1}{T}\sum_{i=1}^T(\hat{\Omega}_{X}(t_i)-\hat{\Omega}^{m}_{X}(t_i))^2$ (with $\hat{\Omega}^{m}_{X}(t_i)=\frac{1}{T}\sum_{i=1}^T\hat{\Omega}_{X}(t_i)$), where $\Omega_{X}(t_i)$ and $\hat{\Omega}_{X}(t_i)$ denote the theoretical and estimated values (mean over $R$ realizations) of the OIR at a specific time instant.
The entire process of generation of time series and estimation of the OIR was repeated 100 times to increase the robustness of the successive statistical analysis. Moreover, after averaging ${\Omega}_{X}(t_i)$ across 100 iterations, the fall time ($F_t$), defined as the time required for the OIR estimate to decrease from 90\% of its highest value to 10\% of its lowest value, was also computed. 

Fig. \ref{fig3} shows the distributions of $\hat{\Omega}_{X}(t_n)$ across 100 iterations, computed for different values of $1-c$ and numbers of realizations $R$, and compared with the exact values derived from the true model parameters (red line). The results underscore the significant influence of the number of realizations and the forgetting factor on the bias-variance trade-off and on the time required to detect transitions. Specifically, the number of realizations influences both the bias and the variance of the estimates, which are reduced substantially moving from $R=10$ to $R=50$.
On the other hand, the forgetting factor appears to affect the bias and variance of the TV-OIR estimates in a more complex way: while the variance is always reduced at increasing of the forgetting factor, the bias is affected in a way that also depends on the waveform whereby the coupling coefficient changes. In fact, the bias tends to decrease at increasing the forgetting factor, but for high values of the latter the bias is influenced also through the response time to transitions, with higher values of $(1-c)$ determining lower responsiveness; the issue is more serious in the case of very fast transitions (square waveform of the coupling coefficient).

A more systematic evaluation of the impact of the number of realizations and the forgetting factor on the TV-OIR estimates was performed using statistical testing. Specifically, a two-way repeated measures ANOVA was carried out separately for the performance measures $\mathrm{BIAS_N}$ and VAR to assess the effects of the factors $R$ and $(1-c)$ on $\hat{\Omega}_{X}(t_n)$. The Mauchly's test of sphericity was applied, and when necessary, the Greenhouse-Geisser correction was used to adjust for violations of the sphericity assumption in all analyses. Post-hoc comparisons between sub-levels of the ANOVA factors were conducted using Tukey’s test. Results are presented in terms of F-values and partial eta squared ($\eta_p^s$) measuring the effect size and summarized in Table \ref{TAB1}. 
The analysis highlights a significant statistical influence of the within-subject factors $R$ and $(1-c)$, and of their interaction $R \times (1-c)$, on both performance measures.

\begin{figure}
\includegraphics{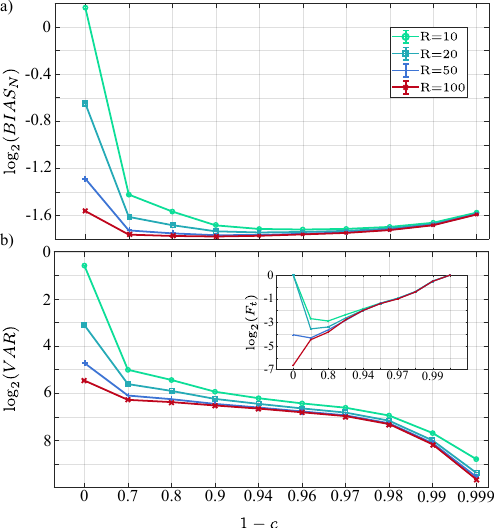}
\caption{Performance measures (a) of bias ($\mathrm{BIAS_N}$) and (b) variance (VAR) as a function of the factors $R$ and $1-c$. Both measures are reported as mean value and 95\% confidence interval computed across 100 iterations for each experimental condition. The inset in (b) reports the fall time computed after averaging $\hat{\Omega}_{X}$ across the 100 iterations.}
\label{fig4}
\end{figure}

Fig. \ref{fig4} shows the distributions of the performance measures as a function of the factors $R$ and $(1-c)$.
The estimation bias ($\mathrm{BIAS_N}$, Fig. \ref{fig4}a) decreases as the number of realizations increases, with an effect that is statistically significant  only when $(1-c) < 0.94$ (Tukey's post-hoc test).
The estimation variance (VAR, Fig. \ref{fig4}b) displays a similar trend, decreasing significantly with increasing the number of realizations $R$ when $(1-c) < 0.94$ but also when $(1-c) > 0.98$. The response time to transitions ($F_t$) reveals a significant impact of the number of realizations $R$ on the RLS algorithm's ability to detect transitions for all values of the forgetting factor $(1-c)\in[0,0.8]$; for instance, when $(1-c)=0$ and $R \in {\{10, 20\}}$, the bias is so high that the fall time reaches its maximum value of 2 seconds, indicating that the transition cannot be detected.

Overall, these results demonstrate that the RLS algorithm enables the time-varying analysis of HOIs in a network of interacting processes. Our findings align with previous studies, which have highlighted the consistency and accuracy of the RLS algorithm in estimating time-varying information-theoretic measures \cite{antonacci2023time,antonacci2024exploring}, as well as coupling and causality measures \cite{moller2003fitting,moller2001instantaneous,astolfi2008tracking,milde2010new}. We find that the bias$/$variance trade-off is more influenced by the number of realizations than by the value of the forgetting factor. Nevertheless, the forgetting factor directly affects the adaptation speed to transitions, which remains largely independent of the number of realizations, as also shown in \cite{moller2003fitting}. Notably, when a very large number of realizations is available the forgetting factor can be set to very low values, allowing for precise identification of fast transitions occurring in the HOIs of the observed dynamic system. If this is not the case, the analysis conducted here identifies a suitable range for the forgetting factor, $0.96 \leq (1-c) \leq 0.98$, to achieve an optimal balance between bias, variance, and the ability to track abrupt transitions, as also highlighted in a previous work \cite{antonacci2023time}.
To provide practical recommendations we advise that, if the analysis aims to determine the time of transition in a system with minimal uncertainty when a sufficient number of realizations is available (e.g., $R > 50$), a small value for the forgetting factor is recommended. Conversely, if the number of realizations is limited (e.g., $R \leq 20$), a value of $(1-c)$ around 0.96-0.98 is recommended to allow proper TV-VAR identification. It is important to note that the VAR model identified at each time step $t_n$ (when $1-c=0$) via the OLS method cannot ensure a stable solution when the number of data samples available for the estimation procedure ($R$ in this case) is at least one order of magnitude greater than the number of VAR parameters to be estimated \cite{antonacci2020information,antonacci2024measuring}.

\begin{table}
\renewcommand{\arraystretch}{1.3}
\caption{Two-way ANOVA (Degrees of
Freedom (DoF) and F values with effect size ($\eta_p^2$) ) computed considering $\text{BIAS}_N$ and VAR as dependent variables; **,
$p<10^{-5}$}. \ \\
\centering
\begin{tabular}{l l l l l}
    \hline
    \hline
    Factor  &  DoF & $\text{BIAS}_N$ ($\eta^2_p$) & VAR ($\eta^2_p$) \\
    \hline
    R & 3, 297 & 3355$^{**}$ (0.971) & 16900$^{**}$ (0.994)\\
    (1-c) & 9, 891 & 66300$^{**}$ (0.998) & 28200$^{**}$ (0.996) \\
    $\text{R} \times \text{(1-c)}$ & 27, 2673 &28000$^{**}$ (0.996) & 15500$^{**}$ (0.993)\\
    \hline
    \hline
\end{tabular}
\label{TAB1}
\end{table}

\section{Application to Time-Varying Brain Dynamics}
This section reports the application of the proposed framework for the time- and frequency-specific analysis of HOIs to benchmark epicranial EEG signals recorded during whisker stimulation in rats \cite{quairiaux2010functional,quairiaux2011functional}. 
The sensorimotor system of rodents represents a simple model of large-scale networks where the structural and functional pathways are well known and established. In this specific context, the brain regions activated and the timing of stimulus encoding have been previously reported, together with analyses investigating causal brain interactions and information dynamics \cite{quairiaux2011functional,plomp2014physiological,pagnotta2018time,antonacci2024exploring}.
Here, we analyze the spatiotemporal and spectral patterns of the OIR with the aim of characterizing the extent, latency and frequency bands of the high-order mechanisms involved in brain information processing.

\subsection{Experimental protocol and data analysis}

\begin{figure}
\includegraphics[scale=0.95]{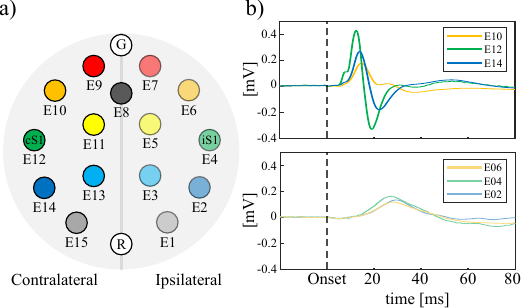}
\caption{(a) Electrode montage for epicranial multichannel EEG. (b) Grand-average SEP for the contralateral (top) and ipsilateral (bottom) hemisphere to the right whisker stimulation.}
\label{fig5}
\end{figure}

The analyzed dataset refers to epicranial EEG data recorded from 10 young Wistar rats according to the procedures detailed in previous works \cite{quairiaux2010functional,quairiaux2011functional}. Specimens were anesthetized with light isoflurane maintained at 2.5\% and mounted in a stereotaxic frame providing a continuous flow of isoflurane in the same air mixture. Epicranial signals were recorded with a custom-made amplifier from an array of 16 stainless steel electrodes 500$\mu m$ in diameter placed to cover the entire skull surface without touching head muscles (gain 5000$\times$; band-pass filters 1-500 Hz; final impedance  $\sim 50 k\Omega$). Electrode coordinates from bregma for the right and left hemispheres
were (rostrocaudal/mediolateral) as follows (Fig. \ref{fig5}a): –7.5/4 mm (E1, E15), –4.75/5 (E2, E14), –3.5/2.25 (E3, E13), –1.5/5 (E4, E12), -0.75/2.25 (E5, E11), 1.25/4 (E6, E10), 3.25/2.25 (E7, E9), 0/0 (E8), –6.25/0 (R), and ground electrode at 6/0 (G). All differential voltages were digitally converted at 2 kHz using custom-made scripts. 

Unilateral stimuli were delivered simultaneously to all large whiskers on one side of the snout through a solenoid-based stimulator device. Large whiskers on one side of the snout were glued together and inserted in a thin tube attached to the stimulator probe that was then placed 1 cm away from the whisker pad. Each stimulus consisted of 500 $\mu m$ backand-forth deflections with 1 msec rise time. Right-sided series of 50 stimuli were applied with an interstimulus interval of 9 s. The entire dataset can be found at \href{https://figshare.com/articles/dataset/epicranialEEG_10rats_zip/5909122/1}{https://doi.
org/10.6084/m9.figshare.5909122.v1}.

We firstly examined the somatosensory evoked potentials (SEPs) obtained through a straightforward averaging procedure, focusing on the activity recorded at specific electrodes in response to the 50 stimuli delivered. The grand-average analysis of SEPs , illustrated in Fig. \ref{fig5}b, reveals a voltage peak over the primary sensory cortex contralateral (cS1, electrode E12) initiating  $\sim 5$ msec after the onset of stimulation and vanishing $2\sim 5$ msec later \cite{quairiaux2011functional,plomp2014physiological}. Additionally, the SEP cS1 has well-defined structural connections to particular contralateral parietal and frontal sensory-motor regions (electrodes E14, E10), which become active immediately following cS1 (Fig. \ref{fig5}b, upper panel). Moreover, the ipsilateral S1 (iS1, eletrode E4) and its neighbor electrodes (E2, E6) display an involvement confined to late latencies; as suggested in previous works \cite{wiest2005heterogeneous} (Fig. \ref{fig5}b, lower panel), although this secondary network shares the same structural and functional connections as the contralateral hemisphere, these connections are less pronounced. 

\subsection{Time-frequency OIR analysis}
To analyze the time-frequency modifications of possible high-order behaviors emerging during whisker stimulation, we considered the EEG signals recorded from six electrodes placed in the contralateral (E10, E12, E14) and ipsilateral (E2, E4, E6) brain hemispheres.
For each specimen, the signals recorded from these electrodes during repeated stimuli were interpreted as realizations ($R=50$) lasting 180 msec (from 60 msec before to 120 msec after the stimulus; $T=360$ samples, $f_s=2000$ Hz) of a vector stochastic process $Y$ composed by $M=6$ scalar processes, i.e., $Y_1=E2, Y_2=E4, Y_3=E6, Y_4=E10, Y_5=E12, Y_6=E14$. 
The analysis was then performed computing the time-frequency OIR for all multiplets of order 3 (20 triplets), order 4 (15 multiplets), and order 5 (6 multiplets), as well as for the multiplet including all 6 processes.
For each specimen, the OIR measures were computed after identifying the TV-VAR model (\ref{TV-VAR}) as described in Sect. \ref{time-frequency HOIs},\ref{identification}.
Identification was performed on the six time series reduced to zero mean and unit variance
optimizing the model order via the MSPE procedure, setting the forgetting factor to $(1-c)=0.975$ \cite{antonacci2023time} and the order $q$ of the reduced models to 30 \cite{sparacino2023statistical}, and initializing the coefficient matrix ($\mathbf{A}(t_p)$) with coefficients randomly sampled from a uniform distribution in the range $[-1,1]$.

\begin{figure}[t!]
\includegraphics{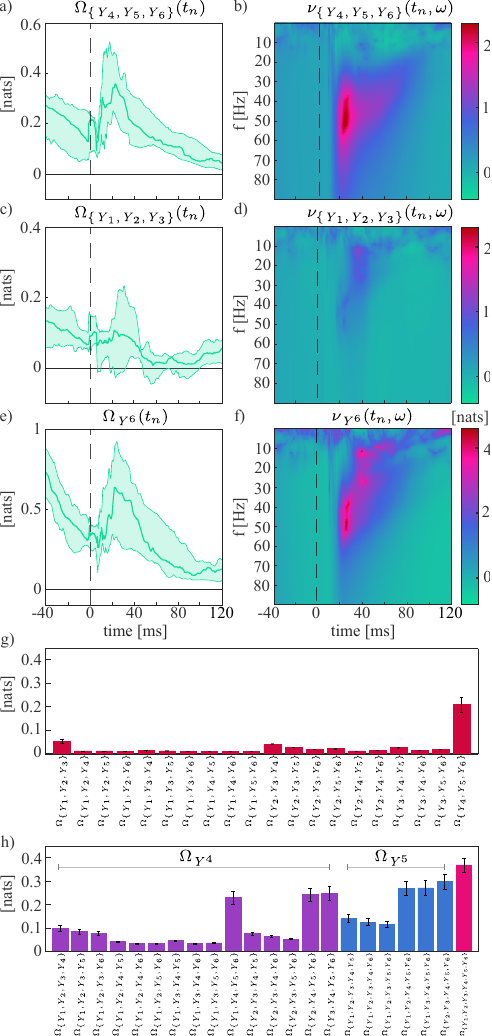}
\caption{Time-resolved and time-frequency analysis of OIR in rats during whisker stimulation.  (a,c,e)The time-varying and (b,d,f) time-frequency profiles of the OIR (median and quartiles across specimens) are reported for the triplets of signals recorded in the (a,b) contralateral and (c,d) ipsilateral hemispheres as well as (e,f) for the multiplet including all six signals. (g, h) Distribution (median and quartiles across specimens) of the OIR values computed for all multiplets of (g) order 3 and (h) orders 4,5,6 obtained integrating the time-frequency OIR over the entire frequency spectrum, and taking the average over the whole duration of the analyzed time window.}
\label{fig6}
\end{figure}

The results are presented in Fig. \ref{fig6}, showing the time-varying and time-frequency OIR computed for the triplets of signals collected in the two hemispheres and the multiplet including all signals, as well as the overall time-domain values obtained for all possible multiplets.
The time-varying analysis reveals the presence of a redundant interaction of order 3 among the signals recorded in the contralateral hemisphere ($\Omega_{\{Y_4,Y_5,Y_6\}}(t_n,\omega)>0$, Fig. \ref{fig6}a), with an evident peak after $\sim$20 msec from the onset of stimulation; this trend is in line with the analysis of SEPs of Fig. \ref{fig5}. On the other hand, the time-varying OIR relevant to the signals recorded in the ipsilateral hemisphere ($\Omega_{\{Y_1,Y_2,Y_3\}}(t_n,\omega)$, Fig. \ref{fig6}c) highlights a different scenario, with lower values of the OIR without a clear peak. 
These trends are confirmed by the time-frequency analysis (Fig. \ref{fig6}b,d), which also allows to localize the frequency bands where redundancy emerges. In the contralateral hemisphere (Fig. \ref{fig6}b), the clear peak of redundancy arising at $\sim$ 20 msec from the stimulation onset is localized in the gamma frequency band (40$\pm$90 Hz), which is physiologically plausible as confirmed by previous studies evidencing the importance of the gamma band for the cS1 \cite{plomp2014physiological,pagnotta2018time}. A slight redundant contribution is also noticeable in the ipsilateral hemisphere (Fig. \ref{fig6}d), occurring at later latencies ($\sim$40-50 msec after the onset) and at lower frequencies, confirming the driving role of iS1 at middle latencies as suggested in \cite{plomp2014physiological}. 
The analysis extended to all 6 signals (Fig. \ref{fig6}e,f) subsumes the trends observed for the triplets, though with higher redundancy values, demonstrating high-order behaviors which originate from the lower order interactions. Specifically, a rise of redundancy is noticed at $\sim 20$ msec from the stimulus in the gamma band, and at $40-60$ msec in the lower beta and alpha bands.
The analysis of the distributions across animals of the OIR computed for all possible multiplets of signals (Fig. \ref{fig6}g,h) documents an increment of redundancy with the order of the analyzed multiplet which is coherent with previous studies analyzing overall time-domain HOIs with time-invariant approaches\cite{sparacino2023statistical,AntonacciSpectral24}. The patterns of HOIs confirm the presence of net redundancy in all the analyzed multiplets, and identify the multiplets including the processes $Y_4$, $Y_5$ and $Y_6$ as those mostly determining the prevalence of redundancy. This finding highlights the dominant role of the contralateral sensory cortex (electrodes E10, E12, E14) in mediating redundant interactions.

Overall, the redundancy-dominated networks observed in both the contralateral and ipsilateral hemispheres can be explained in terms of several previous findings \cite{quairiaux2010functional,quairiaux2011functional,pagnotta2018time,plomp2014physiological} and considering the results from the simulation study performed herein. Specifically, whisker-evoked activity originates in cS1 (process $Y_5$, electrode $E12$), with significant functional outflow occurring at early latencies \cite{quairiaux2011functional}. The functional connections from cS1 preferentially target frontal sensorimotor and parietal regions ($Y_4=E10$ and $Y_6=E14$ in Fig. \ref{fig5}), consistently with the strong structural connectivity between cS1 and these regions \cite{quairiaux2010functional}. These findings also align with the sequential activation patterns observed in the  analysis of SEPs (Fig. \ref{fig5}) \cite{plomp2014physiological,pagnotta2018time}, as well as prior work showing a high degree of causal connectivity among the signals recorded from contralateral electrodes \cite{pagnotta2018time,plomp2014physiological}. Furthermore, our simulation study supports the presence of redundancy originating from a common driver structure, where cS1 influences the dynamics of multiple targets. The time-frequency analysis of OIR further confirms that this interaction is primarily redundant, supporting the hypothesis that redundancy in brain networks enables overlapping information storage across regions, thereby contributing to the stability and resilience of cognitive and motor functions. These findings reinforce the central role of cS1 in coordinating activity across frontal and parietal areas.

\section{Conclusions}
The novel methodology presented in this work significantly advances the set of tools for the study of high-order interdependencies in dynamic network systems. Indeed, while being solidly grounded in frameworks recently introduced to characterize synergy- and redundancy-dominated circuits in networks of random variables \cite{rosas2019quantifying} and random processes \cite{faes2022new},  the proposed time-varying and time-frequency measure enables the analysis of non-stationary and frequency-specific behaviors which are very common in computational neuroscience and physiology.

We demonstrate that the time-resolved and time-frequency analyses are tightly interconnected through the spectral integration property of information-theoretic measures, enabling a consistent derivation of time-varying measures from the expanded time-frequency metric. Our  framework allows for the straightforward computation of measures representing high-order interactions at any order, starting from the identification of time-varying linear models using a recursive least squares estimator with a forgetting factor, which ensures a reliable estimation of both slow transitions and fast intermittent behaviors.

The theoretical example and the application to EEG brain signals showcase the potential of extending information-theoretic measures to capture the balance between redundancy and synergy among large groups of nodes, even in the presence of significant non-stationarities. Moreover, our approach underscores the importance of analyzing interactions in the time-frequency domain, enabling the detection of frequency-specific interactions that may remain hidden in time-domain and even time-varying analyses. The data-driven nature of the proposed method makes it applicable not only to biomedical time series but also to a wide range of dynamic systems, including electronic, climatological, social, and financial networks, where the activity of each node can be modeled and studied using the theory of stochastic processes.

\section{acknowledgments}
Y.A. and L.F. were partially supported by SiciliAn MicronanOTecH Research And Innovation CEnter "SAMOTHRACE" (MUR, PNRR-M4C2, ECS$\_$00000022), spoke 3 - Università degli Studi di Palermo "S2-COMMs - Micro and Nanotechnologies for Smart $\&$ Sustainable Communities. L.F. was partially supported by PRIN 2022 project “HONEST-High-Order Dynamical Networks in Computational Neuroscience and Physiology: an Information-Theoretic Framework” (funded by MUR, code 2022YMHNPY, CUP B53D23003020006). L.M. gratefully acknowledges the support of the "Hundred Talents" program of the University of Electronic Science and Technology of China, of the "Outstanding Young Talents Program (Overseas)" program of the National Natural Science Foundation of China, and of the talent programs of the Sichuan province and Chengdu municipality.

\nocite{*}

\bibliography{biblio}

\end{document}